\documentclass[aps,prd,twocolumn,floatfix,showpacs,superscriptaddress,nofootinbib]{revtex4-2}
\usepackage{amsmath,amsfonts,amssymb,bm}
\usepackage{diagbox}
\usepackage{graphicx}
\usepackage{xcolor}
\usepackage{subfigure}
\usepackage{multirow}
\usepackage{textcomp}
\usepackage{slashed}
\usepackage[bottom]{footmisc}
\usepackage[T1]{fontenc}
\definecolor{purple}{rgb}{0.5,0,0.5}
\definecolor{blue}{rgb}{0.0,0,1.0}
\usepackage[colorlinks=true, pdfstartview=FitV, linkcolor=purple, citecolor= purple, urlcolor=blue]{hyperref}

\newcommand{\beq}{\begin{equation}}
\newcommand{\eeq}{\end{equation}}
\newcommand{\ba}{\begin{array}}
\newcommand{\ea}{\end{array}}
\newcommand{\bea}{\begin{align}}
\newcommand{\eea}{\end{align}}
\newcommand{\bi}{\begin{itemize}}
\newcommand{\ei}{\end{itemize}}
\newcommand{\ben}{\begin{enumerate}}
\newcommand{\een}{\end{enumerate}}
\newcommand{\bc}{\begin{center}}
\newcommand{\ec}{\end{center}}
\newcommand{\bl}{\begin{flushleft}}
\newcommand{\el}{\end{flushleft}}
\newcommand{\br}{\begin{flushright}}
\newcommand{\er}{\end{flushright}}

\newcommand\comment[1]{ \hbox{[{\it Comment suppressed here.}\/]} }
\newcommand\hide[1]{}
\newcommand{\skipover}[1]{}

\begin{document}
\title{Isospin breaking in $\rho\pi\pi$ couplings: $\rho^{+,0}$ width difference and muon $g-2$ }

\author{A. Bashir}%
\email{adnan.bashir@dci.uhu.es}
\affiliation{Department of Integrated Sciences and Center for Advanced Studies in Physics, Mathematics and Computation, University of Huelva, E-21071, Huelva, Spain.}
\affiliation{Tecnológico Nacional de México/Instituto Tecnológico de Morelia, División de Estudios de Posgrado e Investigación, Av. Tecnológico 1500, Lomas de Santiaguito, Morelia CP 58120, M\'exico.}

\author{G. L\'opez Castro}%
\email{gabriel.lopez@cinvestav.mx}
\affiliation{Departamento de Física, Centro de Investigación y de Estudios Avanzados del IPN, Apartado Postal 14-740, CP 07000 Ciudad de M\'exico, M\'exico.}

\author{K. Raya}
\email{khepani.raya@dci.uhu.es}
\affiliation{Department of Integrated Sciences and Center for Advanced Studies in Physics, Mathematics and Computation, University of Huelva, E-21071, Huelva, Spain.}
\affiliation{Departamento de Actuaría, Física y Matemáticas, Universidad de las Américas Puebla, San Andrés Cholula, 72820. Puebla, México.}

\date{\today}

\begin{abstract}
The dominant contribution to the hadronic vaccum polarization (HVP) of the muon  $g-2$ can be calculated using the hadronic spectrum in $\tau^- \to \pi^-\pi^0\nu_{\tau}$ decay, provided isospin breaking (IB) corrections are properly accounted for. Among these corrections, IB in the $\rho\pi\pi$ coupling has been pointed out as a missing ingredient. In this work, we calculate this IB effect within the impulse approximation using a symmetry-preserving
 framework based on
Dyson-Schwinger and Bethe-Salpeter equations, incorporating fully dressed quark propagators and meson Bethe-Salpeter amplitudes.  We then estimate its effect in the HVP contribution to the muon $g-2$. As a natural extension, we also compute the IB in the ratio of $\phi K^+K^-/\phi K^0\bar{K^0}$ couplings and compare with existing data.
 We obtain a coupling ratio
$g_{\rho^0\pi^+\pi^-}/g_{\rho^+\pi^+\pi^0}
= 1.0024(17)(1)$,
which yields an isospin-breaking correction of
$\Delta a_\mu^{\mathrm{HVP,LO}}[\rho\pi\pi]
= -2.32(1.64)\times 10^{-10}$
to the two-pion contribution to the muon anomalous
magnetic moment extracted from $\tau$ decay data.

\end{abstract}

\maketitle

\noindent
\section{Introduction}
Isospin is an approximate, although  very good symmetry of strong interactions owing to its breaking by the mass and charge difference of $u$ and $d$ quarks. Isospin symmetry breaking (IB) effects in observables like mass splitting within isospin multiplets, the nuclear force, or hadronic reactions, turn out to manifest below the few percent level \cite{ParticleDataGroup:2026mpi}. Precision calculations of IB are generally limited by the model-dependency (except perhaps in lattice QCD), although they are required as inputs in precise tests of the Standard Model (SM)  as in the prediction of the muon $g-2$ \cite{Aliberti:2025beg}, or determinations of quark mixing angles like $|V_{ud}|, |V_{us}|$ \cite{Hardy:2020qwl,ParticleDataGroup:2026mpi}. 

Isospin breaking in the strong $VPP$ (where $V/P$ is a vector/pseudoscalar meson) couplings is relevant to predict the dominant hadronic vacuum polarization (HVP) contribution to the muon $g-2$ using tau data \cite{Aliberti:2025beg}. The  production rates of charged/neutral  $K\bar{K}/D\bar{D}/B\bar{B}$ pairs in $e^+e^-$ colliders at the   $\phi/\psi(2S)/\Upsilon(4S)$ resonances, which are relevant to determine the number of these flavored mesons, depend critically of isospin breaking as they occur close to threshold. While IB of electromagnetic origin have been calculated in the form of radiative corrections to the $VPP$ vertices \cite{Cremmer:1969er, Lepage:1990hm, Byers:1990rd,Kaiser:2002bm, Voloshin:2003gm, Flores-Baez:2008owd}, effects due to the $u-d$ mass difference are missing in most cases.

Our main focus in this paper is the IB in the ratio of $\rho^+\pi^+\pi^0/\rho^0\pi^+\pi^-$ couplings, which is relevant in the calculation of the width difference of $\rho(770)$ vector mesons $\Delta \Gamma_{\rho}=\Gamma_{\rho^0}-\Gamma_{\rho^+}$. As pointed out in Ref. \cite{Aliberti:2025beg}, our ignorance on this IB effect may be the largest source of uncertainty in the prediction of the leading order HVP contribution to the muon $g-2$ calculated with $\tau$ lepton data. This missing piece of information forced to adopt in Ref. \cite{Aliberti:2025beg} an uncertainty in IB  couplings of order $\delta g/g\equiv (g_{\rho^+\pi^+\pi^0}/g_{\rho^0\pi^+\pi^-}-1)=\pm 2\alpha/\pi$. This potential uncertainty induces a rather large uncertainty $\pm 4.7\times 10^{-10}$ in the isospin breaking correction contributions to the HVP of  muon $g-2$ \cite{Aliberti:2025beg}. Clearly, the uncertainty in the calculation of $\delta g/g$ must be at least three times smaller to be competitive with the current experimental uncertainty \cite{Muong-2:2025xyk}, which also motivates the present study.

We explore IB effects in the strong decay $\rho\to\pi\pi$ within the Dyson-Schwinger (DS) and Bethe-Salpeter (BS) equations framework. The former are the Euler-Lagrange equations in a quantum field theory, and the latter provide a natural Poincaré-covariant description of the meson's internal structure\,\cite{Roberts:1994dr,Eichmann:2016yit,Raya:2024ejx}. In principle, the full dynamics is encoded in the infinite tower of coupled DS equations, whose solutions enter the BS equation. Hence, the quark self-energy and bound-state interaction kernels are related through symmetry principles, establishing a mutual interplay between the dressing of the quark propagator and the dynamics of the bound state. In practice, the infinite tower of DS equations must be systematically truncated to obtain a tractable description\,\cite{Chang:2009zb,Qin:2020rad,Qin:2020jig}. Rather than pursuing a numerically involved approach, herein we adopt the symmetry-preserving contact-interaction treatment described in Refs.\,\cite{Xing:2022jtt,Xing:2021dwe,Wang:2022mrh,Dang:2023ysl,Xing:2024bpj,Xu:2024frc,Xing:2025iab,Xu:2026zbs}, which extends traditional approaches. This framework provides, for the first time, a symmetry-preserving setting in which to explore IB effects in the $g_{\rho^0\pi^+\pi^-}$ and $g_{\rho^+\pi^+\pi^0}$ couplings. The procedure is described below. As an application of the IB in $\rho\pi\pi$ couplings we calculate the width difference of $\rho^+-\rho^0$ mesons and evaluate the corresponding IB effects in the muon $g-2$ calculated using tau data.

In addition, we also  calculate  the isospin breaking in $\phi K^+K^-/\phi K^0 \overline{K^0}$ couplings  within the same framework. We find that the calculated effect in this case is compatible with current measurements of the relative production of charged/neutral kaons in $\phi$ meson decays.

\section{Strong decays}

\subsection{Definitions}
The strong couplings $V\to PP'$  can be defined from the on-shell matrix element\,\cite{daSilveira:2022pte,Xu:2024rew,Jarecke:2002xd}:
\begin{equation}
\label{eq:MatElem}
    \langle P'(p_2)P(q)|V(p_1,\lambda)\rangle=g_{V P P'} \epsilon^{\lambda} \cdot q\,,
\end{equation}
where $\epsilon_\mu^\lambda$ is the polarization vector of the $V$ meson, and the on-shell four-momenta are:
\begin{equation}
    p_1^2=-m_V^2\,,\,p_2^2=-m_{P'}^2\,,\,q^2=-m_P^2\,,
\end{equation}
with $m_X$ denoting the corresponding meson mass. The decay amplitude can be evaluated via the so called impulse approximation, which is fully characterized by the dressed quark propagators ($S^f$) and the bound-state BS amplitudes ($\Gamma_X$) as follows:
\begin{eqnarray}\nonumber
    g_{VPP'} \epsilon^\lambda\cdot q&=&N_c\text{tr}\int_k S^{f_3}(k_3)\epsilon^\lambda \cdot\Gamma_{V}(k_V;p_1)S^{f_1}(k_1)\\
    &\times&\bar{\Gamma}_{P'}(k_{P'},-p_2)S^{f_2}(k_2)\Gamma_{P}(k_P,-q)\,.
    \label{eq:defIA}
\end{eqnarray}
Here, $f_i$ are the quark flavors, $\text{tr}$ denotes a trace over Dirac indices, $N_c$ is the number of colors, and $\int_k=\int d^4k/(2\pi)^4$. The momentum flow reads:
\begin{subequations}
\begin{align}
    k_1 &= k + w_1 p_1 \,, \\
    k_2 &= k + w_1 p_1 - p_2 \,, \\
    k_3 &= k - w_2 p_1 \,.
\end{align}
\end{subequations}
such that $w_1+w_2=1$ defines de momentum partition. The relative BS amplitude momenta are:
\begin{subequations}
\begin{align}
    k_V &= k + \frac{1}{2}(w_1-w_2)p_1 \,, \\
    k_{P'} &= k + w_1 p_1 - \frac{1}{2}p_2 \,, \\
    k_P &= k + \frac{1}{2}(w_1-w_2)p_1 - \frac{1}{2}p_2 \,.
\end{align}
\end{subequations}

\begin{figure}[htbp]
    \centering
    \includegraphics[width=0.65\columnwidth]{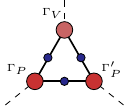}
    \caption{Pictorial representation of the impulse approximation employed to compute the $g_{VPP'}$ strong couplings, Eq.\,\eqref{eq:defIA}. The colored blobs represent dressed propagators and BS amplitudes.}
    \label{fig:triangle}
\end{figure}

For illustration purposes, the triangle diagram associated with Eq.\,\eqref{eq:defIA} is shown in Fig.\,\ref{fig:triangle}. The computation of all elements entering this equation is described below.

\subsection{One-body and two-body kernels}
The quark DS equation, commonly referred to as the gap equation, describes the dynamical dressing of the quark through its self-interactions, thereby encoding both the perturbative and nonperturbative dynamics of QCD. It is expressed as follows\,\cite{Sultan:2018tet,Serna:2018dwk,Chang:2021vvx}:
\begin{equation}
\label{eq:quarkPropGen}
    S^{-1}(p)=[S^{(0)}(p)]^{-1}+\int_{q} \mathcal{K}^{(1)}(q,p)S(q)\;,
\end{equation}
where $S^{(0)}(p)=[i \gamma \cdot p+m^{\text{bm}}]^{-1}$ corresponds to the bare quark propagator, with a Lagrangian current quark mass $m^{\text{bm}}$, and $\mathcal{K}^{(1)}(q,p)$ the one-body kernel. For convenience, we have supressed all color and flavor indices, as well as renormalization constants. 

\begin{figure}[htbp]
    \centering
    \includegraphics[width=0.95\columnwidth]{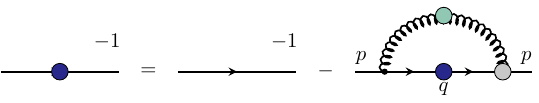}
    \caption{Quark gap equation as defined in Eqs.\,\eqref{eq:quarkPropGen}-\,\eqref{eq:K1}. Dressed quantities, namely propagators and vertices, are represented with colored blobs.}
    \label{fig:DSE}
\end{figure}

The one-body kernel $\mathcal{K}^{(1)}(q,p)$  takes the form:
\begin{equation}
\label{eq:K1}
    \mathcal{K}^{(1)}(q,p)=\frac{4}{3} g^2 D_{\mu\nu}(p-q)\gamma_\mu \otimes \Gamma_\nu(q,p)\;,
\end{equation}
where $g$ is the Lagrangian coupling constant; $D_{\mu\nu}$  and $\Gamma_\nu$ are the fully-dressed gluon propagator and quark-gluon vertex, respectively. Each of these components obey their own DS equation, depending on higher-order Green functions, thus yielding an infinite set of coupled equations. A pictorial representation of the quark DS equation can be found in Fig.\,\ref{fig:DSE}. 

In analogy to its bare structure, the fully-dressed quark propagator can be generically expressed as:
\begin{equation}
\label{eq:quarkpropdef}
    S^{-1}(p) =Z(p^2)[i\gamma \cdot p + M(p^2)]\,,
\end{equation}
so that the scalar functions $Z(p^2)$ and $M(p^2)$ capture the effects of quark self-interaction beyond the tree-level. In particular, the infrared enhancement of the quark mass function $M(p^2)$, provides a direct manifestation of dynamical chiral symmetry breaking (DCSB), whose relative impact depends on the quark flavour and its associated current-quark mass.

The properties of valence quark/antiquark bound-states are encoded within solutions of its Poincaré covariant BS equation\,\cite{Roberts:1994dr,Eichmann:2016yit,Raya:2024ejx}:
\begin{equation}
\label{eq:BSEGen}
    \Gamma_{X}(p;P)=\int_{q} \mathcal{K}^{(2)}(q,p;P)\chi_{X}(q;P)\;,
\end{equation}
where $\chi_{X}(q;P)=S^{f_1}(q_+)\Gamma_{X}(q;P)S^{f_2}(q_-)$ denotes the BS wavefunction. For convenience we have defined:
\begin{equation}
q_+=q+\eta P\,,\,q_-=q-(1-\eta)P\,,
\end{equation}
where $\eta\in(0,1)$ controls the momentum partition parameter. In any Poincaré-covariant framework, the result is independent of it. The two-particle irreducible quark/antiquark scattering kernel is denoted by $\mathcal{K}^{(2)}(q,p;P)$. Finally, the BS amplitude is defined by a specific combination of Dirac matrices that specify the $J^{PC}$ channel. Below we present the specific structures that emerge within our scheme.

Evidently, the quark propagator enters the BS equation, thereby establishing a feedback between the quark self-interaction and the bound-state dynamics. To decouple these equations from the rest of the tower, a systematic truncation scheme is required. The vector and axial-vector Ward-Green-Takahashi identities (WGTIs) provide the proper means to achieve it. In the chiral limit, these identitites read\,\cite{Chang:2009zb,Qin:2020rad,Qin:2020jig}:
\begin{subequations}\label{eq:WGTIs}
\begin{align}
i P_{\mu}\Gamma_{\mu}(k;P)
&= S^{-1}(k_{+})-S^{-1}(k_{-}) \,,
\label{eq:vWGTI}\\
P_{\mu}\Gamma_{5\mu}(k;P)
&= S^{-1}(k_{+})i\gamma_{5}
  +i\gamma_{5}S^{-1}(k_{-}) \,,
\label{eq:avWGTI}
\end{align}
\end{subequations}
where $\Gamma_{\mu}$ and $\Gamma_{5\mu}$ are the vector and axial-vector vertices. The above establishes a close connection between the one-body quark self-energy and the two-body BS kernel. Thus, modifications of the former, such as those induced by explicit IB (namely, the $m_d/m_u$ ratio), must be consistently reflected in the latter. At the same time, the BS kernel can itself generate dynamical IB effects via interactions within the bound-sate. Furthermore, WGTIs enforce crucial symmetries, such as the emergence of the pion as the Nambu-Goldstone boson of DCSB and the vanishing of the decay constants of its radial excitations in the chiral limit\,\cite{Munczek:1994zz,Bender:1996bb,Holl:2005vu}.

For the above, the BS kernel cannot be chosen independently of the gap equation kernel, and viceversa. The simplest symetry-preserving choice is the so-called rainbow-ladder (RL) approximation, defined through:

\begin{subequations}
\label{eq:RL}
\begin{align}
\label{eq:K1RL}
    &\mathcal{K}^{RL}(q,p):=-\frac{4}{3}D_{\mu\nu}^{\text{eff}}(p-q)\gamma_\mu \otimes \gamma_\nu\;,&\\
   & \mathcal{K}^{(1)}(q,p)=-\mathcal{K}^{(2)}(q,p;P)=-\mathcal{K}^{RL}(q,p)\,,&
\end{align}
\end{subequations}
where the gluon propagator $g^2 D_{\mu\nu}$ is demoted to an effective coupling $D_{\mu\nu}^{\text{eff}}$, and the fully dressed quark gluon vertex is reduced to $\Gamma_\nu\to\gamma_\nu$. 

The RL scheme is practical and valuable, enabling a broad range of pseudoscalar meson properties to be described systematically; see e.g. Refs.\,\cite{Ding:2022ows,Qin:2020rad,Raya:2024ejx}. Nevertheless, it has well-kown limitations, particularly in the description of radially and orbitally excited states, e.g.\,\cite{Qin:2011xq,Mojica:2017tvh,Qin:2020jig,Xu:2022kng}. Other drawbacks arise when vector mesons are involved. For instance, calculations of the coupling $g_{\rho\pi\pi}$ consistently yield values $20\%$ smaller than expected\,\cite{daSilveira:2022pte,Xu:2024rew,Jarecke:2002xd}. More importantly for the present work, the RL truncation does not incorporate IB breaking effects. Some ideas have been discussed in Ref.\,\cite{Miramontes:2022mex}, albeit the proposed kernels produce small violations of the WGTIs, whose impact on physical observables is hard to gauge. Aiming at providing the symmetry-preserving character of the DS-BS equations approach, we follow an alternative path. 

It has been shown in e.g. Refs.\,\cite{Xing:2021dwe,Chang:2020iut} that WGTIs do not uniquely determine the structure of the two-body kernel. This freedom allows the kernel to be decomposed as\,\cite{Dai:1990ap}:
\begin{subequations}
\label{eq:BRLdef}
    \begin{align}
        &\mathcal{K}^{(2)}(q,p;P)=\mathcal{K}^{RL}(q,p)+\mathcal{K}^{BRL}(q,p;P)\,,&\\
        &\mathcal{K}^{BRL}(q,p;P):=\tilde{D}^j(q,p) \tilde{\Gamma}_{j}(q,p)\otimes \tilde{\Gamma}_{j}(q,p) F^j(q,p)\;,&
    \end{align}
\end{subequations}
where $F^j$ are Lorentz invariant scalar functions and $\tilde{\Gamma}_{j}$ denote  combinations of Dirac matrices; the term $\tilde{D}^j$ is auxiliary and, when necessary, provides the tensor structures needed to contract the Lorentz indices of $\tilde{\Gamma}_{j}$. As shown in Ref.\,\cite{Xing:2021dwe}, the particular choice
    \begin{equation}
    \label{eq:BRLker1a}
        \tilde{\Gamma}_{j}(q,p) = \{ \mathbb{I},\;\gamma_5,\;i/\sqrt{6}\sigma_{\mu\nu}\}\,,
    \end{equation}
together with $\tilde{D}^j(p,q)F^j(p,q)  \to \xi \tilde{D}$, preserves the symmetry constraints imposed by the WGTIs, Eqs.~\eqref{eq:WGTIs}. Expressed in this way, the dressing of the BRL component, $\xi \tilde{D}$, becomes a momentum-independent quantity, setting the strength of the BRL contribution. 

\begin{figure}[htbp]
    \centering
    \includegraphics[width=0.95\columnwidth]{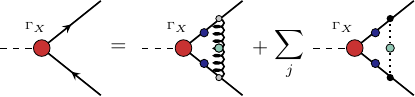}
    \caption{Meson BS equation in the present BRL framework. The first term on the right corresponds to the typical one-gluon exchange of the RL truncation, whereas the second terms encodes the effective exchanges of $j=$ scalar, pseudoscalar and tensor degrees of freedom (see Eqs.\,\eqref{eq:RL}-\eqref{eq:BRLker1a}). Dressed quantities are represented by colored blobs. }
    \label{fig:BSE}
\end{figure}

The BRL piece encodes effective exchanges of massless scalar, pseudoscalar and tensor degrees of freedom in the two-body bound-state equation. An illustration is provided in Fig.\,\ref{fig:BSE}. The BRL prescription has been shown to improve the description of the vector channels, while keeping the pseudoscalar meson properties untouched. Furthermore, promoting $\xi$ to a flavor-dependent quantity, $\xi := \xi_f \xi_g$, causes these effective exchanges to dynamically induce IB effects in the vector sector, producing:
\begin{equation}
    \Delta m_\rho:=m_{\rho^+}-m_{\rho^0}\neq 0\,.
\end{equation}
This is a key characteristic of the present approach.

Before proceeding to the evaluation of the strong $VPP$ couplings, we introduce an additional prescription that substantially reduces the numerical complexity while retaining the nonperturbative character of QCD and preserving the relevant symmetry constraints, the so-called contact interaction (CI) model.

\subsection{Contact interaction model}
Let us consider the gap and BS equations defined in Eq.\,\eqref{eq:quarkPropGen} and Eq.\,\eqref{eq:BSEGen}, respectively, together with the interaction kernels presented through Eqs.\,\eqref{eq:RL}-\eqref{eq:BRLker1a}. The CI model is defined via a momentum-independent gluon propagator:
\begin{equation}
    \label{eq:gluonprop}
    g^2D_{\mu\nu}^{\text{eff}}(p-q) \to \frac{1}{m_G^2}\delta_{\mu\nu}\;,
\end{equation}
where $m_G=0.132\,\text{GeV}$ resembles the effective infrared mass scale generated in the gauge sector of QCD\,\cite{Aguilar:2009nf,Cui:2019dwv,Ferreira:2025anh}. Having absorbed $g^2$ into its definition, we implicitly assumed that the effective gluon also incorporates all dynamical dressing of the quark–gluon vertex. Consequently, $\Gamma_\nu\to\gamma_\nu$, and hence:
\begin{equation}
\label{eq:quarkPropCI}
    S^{-1}(p)=[S^{(0)}(p)]^{-1}+\frac{4}{3 m_G^2} \int_{q}  \gamma_\mu S(q) \gamma_\mu\;.
\end{equation}
For its part, the meson BS equation takes the form:
\begin{equation}
    \label{eq:CIinMRL}
    \Gamma_H(P)=-\frac{4}{3m_{G}^{2}}\int_{q}\left[\gamma_{\mu}\chi_H(P)\gamma_{\mu} - \xi_f \xi_g \tilde{\Gamma}_{j}\chi_H(P)\tilde{\Gamma}_{j}\right]\,.
\end{equation}
The momentum-independent nature of the CI model results in ultraviolet-divergent integrals, requiring an appropriate regularization procedure. We follow the well-established approach from Refs.\,\cite{Xing:2022jtt,Xing:2021dwe,Wang:2022mrh,Dang:2023ysl,Xing:2024bpj,Xu:2024frc,Xing:2025iab,Xu:2026zbs}, based upon a proper-time regularization scheme, wherein:
\begin{equation}
    \frac{1}{s+z}=\int_0^\infty d\tau e^{
    -\tau(s+z)}\to \int_{\tau_{\rm uv}^2}^{\tau_{\rm ir}^2}d\tau e^{-\tau(s+z)}\,.
\end{equation}
The infrared cutoff $\tau_{\rm ir}^{-1}=0.24\,\text{GeV}\sim\Lambda_{\rm QCD}$ guarantees confinement by ensuring the absence of quark production thresholds, while $\tau_{\rm uv}^{-1}=0.905\,\text{GeV}\sim m_p$ provides an ultraviolet cut-off that sets the scale of all dimensioned quantities. More details concerning the regularization procedure can be found in\,\cite{Xing:2022jtt}. Note that the infrared and ultraviolet cutoffs, together with $m_G$, are fixed according to established values in the literature, see Refs.\,\cite{Gutierrez-Guerrero:2019uwa,Yin:2019bxe,Gutierrez-Guerrero:2021rsx,Yin:2021uom,Hernandez-Pinto:2026vrn}, and are therefore not treated as free parameters in this work.

Any sensible regularization scheme yields momentum-independent dressing functions $Z(p^2)\to1$ and $M(p^2)\to M$, such that $M$ plays the role of a constituent-quark mass. Its value reflects both the explicit current-quark mass and the nonperturbative dynamics of QCD associated with DCSB, with the latter providing the dominant contribution for light quarks. Thus, despite $m_d/m_u\sim2$ at the current-quark level, the overwhelming dynamical component drives the dressed-quark masses toward near degeneracy $M_d/M_u\gtrsim 1$,\,\,\cite{Sultan:2018tet,Serna:2018dwk}.

On the other hand, the CI model produces a rather compact structure for the pseudoscalar and vector meson BS amplitudes. Specifically:
\begin{subequations}
\label{eq:BSAs}
    \begin{align}
        &\Gamma_{0^-}(P)=\gamma_5\left[iE_{0^-}(P)+\frac{\gamma \cdot P}{M}F_{0^-}(P) \right]\,,&\\
        &\Gamma_{1^-}(P)=\gamma_{\mu}^TE_{1^-}(P)+\frac{1}{M}\sigma_{\mu\nu}P_\nu F_{1^-}(P)\,, &
    \end{align}
\end{subequations}
with $\gamma_\mu^T=\gamma_\mu-\frac{\gamma \cdot P}{P^2}P_\mu$. It is worth noting that the scalar functions $E_X$ and $F_X$ are independent of the relative momentum between the valence quark/antiquark, depending only on the total hadron momentum $P$. The smallest value of $P^2$ satisfying the bound-state condition, $P^2=-m_X^2$, identifies the ground-state solution. In the computation of physical observables, the corresponding BS amplitudes must be canonically normalized\,\cite{Xing:2021dwe}.

Having established the DS and BS equations, the next step is determining current-quark masses $m_f$ and the parameters $\xi_f$, which control the relative weight of the BRL contribution. These are the only parameters entering our calculations and, once specified, the quark propagators, meson masses, and BS amplitudes can be determined, providing all the necessary to calculate the strong couplings through Eq.\,\eqref{eq:defIA}.

Before proceeding, we highlight several relevant features of the CI model. The kernels entering the quark propagator and meson Bethe–Salpeter equations preserve the relevant WGTIs, while the momentum-independent gluon interaction incorporates an effective gauge-sector mass scale and is sufficiently strong to generate DCSB. Together with symmetry-preserving regularization, these features provide a confinement-compatible description. Hence, despite its simplified momentum dependence, the CI retains the essential nonperturbative physics of low-energy QCD relevant to the strong couplings studied in the present work.

\begin{figure}[!htpb]
    \centering
    \includegraphics[width=\columnwidth]{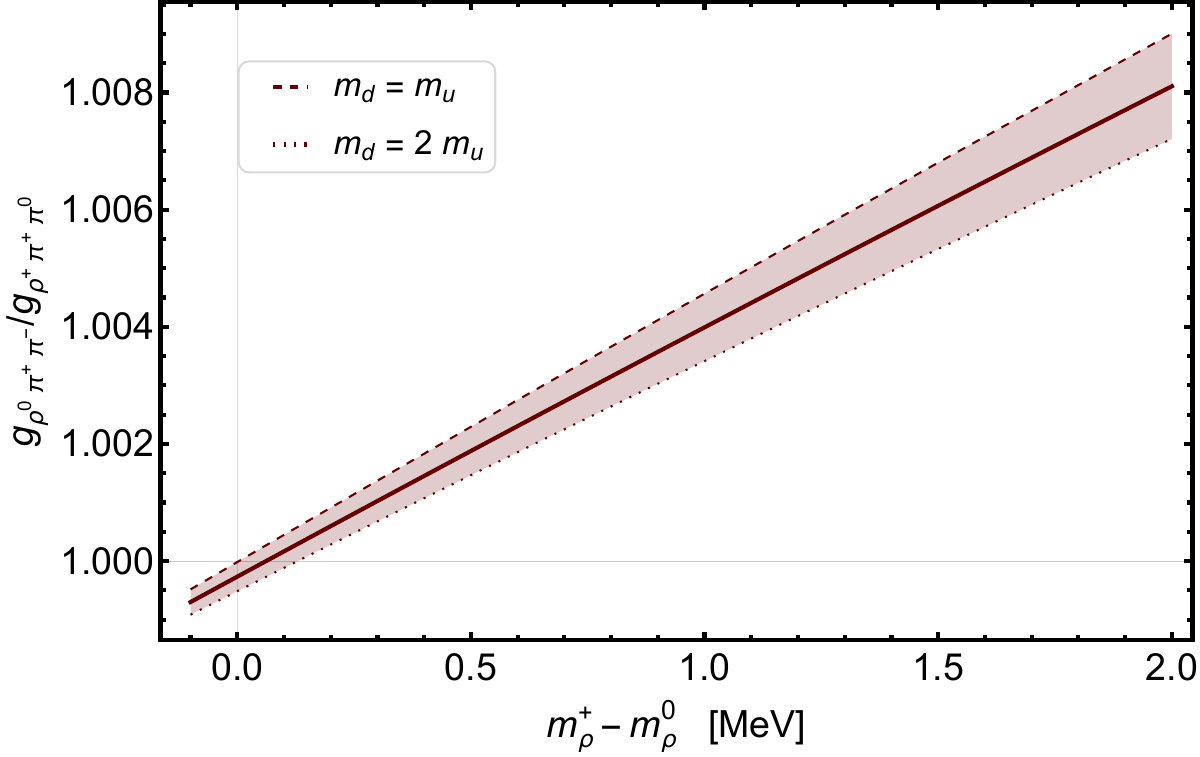}
    \caption{Evolution of the coupling constant ratio $g_{\rho^0\pi\pi}/g_{\rho^+\pi\pi}$ as a function of the mass splitting. The band corresponds to the variation of the ratio $m_d/m_u\in[1,2]$, with $m_\pi=139.47\,\text{MeV}$ fixed.}
    \label{fig:myfigure}
\end{figure}

\section{Isospin breaking in the $g_{\rho\pi\pi}$ couplings}
As discussed above, once the DS and BS have been properly established, the next step is to fix the current-quark masses and the flavor-dependent parameter $\xi_f$, which controls the strength of the BRL contribution, and hence the dynamical IB. The latter has no effect on the pseudoscalar channel and therefore leaves the pion masses degenerate. Nevertheless, its determination is crucial for describing the $\rho$ meson mass-splitting. 

That said, the current-quark masses are fixed by reproducing the ground-state pion mass. Therefore, firstly assuming isospin symmetry $m_u=m_d$, the value $m_{u/d}=m_q=6.98\,\text{MeV}$ produces:
\begin{equation}
\label{eq:mpi}
    m_{\pi^\pm}=m_{\pi^0}=139.47\,\text{MeV}\,,
\end{equation}
in agreement with the empirical charged-pion mass\,\cite{ParticleDataGroup:2026mpi}. Note this value can always be obtained for $m_d/m_u\geq1$ provided $m_u\leq m_q=6.98\,\text{MeV}$, where Eq.\,\eqref{eq:quarkPropCI} yields:
\begin{equation}
    M_u \leq M_q=367.66\,\text{MeV}\,.
\end{equation}

We therefore consider the range $m_d/m_u\in[1,2]$ to assess the uncertainty associated with this explicit source of IB. For all practical purposes, nothing material is gained by fine-tuning the pion mass, or using instead the neutral-pion mass ($m_{\pi^0}\approx135\,\text{MeV}$) or the average value ($m_\pi\approx137.5\,\text{MeV}$) as the benchmark. In fact, the relative impact of this choice on the ratio $g_{\rho^0\pi^+\pi^-}/g_{\rho^+\pi^+\pi^0}$ is at the level of $0.01\%$.

With the constraint imposed by Eq.\,\eqref{eq:mpi}, the seemingly generous range of variation $m_d/m_d$ translates into:
\begin{equation}
\label{eq:MdMu}
    m_d/m_u\in[1,2]\Rightarrow M_d/M_u\in[1,1.02]\,.
\end{equation}
This highlights the dominance of DCSB as the mechanism responsible for mass generation in the quark sector. Moreover, the small variation of the dressed-quark masses favors $g_{\rho^0\pi^+\pi^-}\approx g_{\rho^+\pi^+\pi^0}$, while anticipating that the uncertainty in this ratio will be dominated by dynamical, rather than explicit, IB effects.

Once the current-quark masses have been fixed, the next step is to determine $\xi_u$ and $\xi_d$. In the absence of dynamical IB effects, namely $\xi_u=\xi_d$, the choice $\xi_{u,d}=\xi_q=0.37$ yields:
\begin{equation}
    m_{\rho^\pm}=m_{\rho^0}=775.19\,\text{MeV}\,,
\end{equation}
in line with the experimental average\,\cite{ParticleDataGroup:2026mpi}. The value $m_{\rho^0}$ remains unchanged as long as $\sqrt{\xi_u\xi_d}=\xi_q$. Thus, keeping $\xi_q$ fixed while varying $\xi_u$ modifies $m_{\rho^\pm}$, generating a dynamical mass splitting $\Delta m_\rho$. We therefore vary $\xi_u$ according to: 
\begin{equation}
 \Delta m_\rho\in[-0.1,2]\,\text{MeV}\,,   
\end{equation}
encompassing the current experimental estimate $\Delta m_\rho=0.7(0.8)\,\text{MeV}$ \cite{ParticleDataGroup:2026mpi}. The variation in $\Delta m_\rho$ constitutes the dominant source of uncertainty in the ratio $g_{\rho^0\pi^+\pi^-}/g_{\rho^+\pi^+\pi^0}$.

Thus, for each parameter set $\{m_u,m_d,\xi_u,\xi_d\}$ satisfying the conditions described above, we determine the corresponding quark propagators, meson masses and BS amplitudes, and then use these in Eq.\,\eqref{eq:defIA} to calculate the ratio $g_{\rho^0\pi^+\pi^-}/g_{\rho^+\pi^+\pi^0}$. The dependence of this ratio on $\Delta m_\rho$ and $m_d/m_u$ is illustrated in Fig.\,\ref{fig:myfigure}.

Unsurprisingly, the symmetric limit $m_d/m_u=\xi_d/\xi_u=1$ yields $\Delta m_{\rho}=0$ and, correspondingly, $g_{\rho^0\pi^+\pi^-}/g_{\rho^+\pi^+\pi^0}=1$ (see Fig. \ref{fig:myfigure}). In this scenario\,\footnote{ Relating this quantity to the corresponding decay width, and using the lastest Particle Data Group values,\,\cite{ParticleDataGroup:2026mpi}, $g_{\rho\pi\pi}\approx5.94$.} :
\begin{equation}
    g_{\rho^0\pi^+\pi^-}=g_{\rho^+\pi^+\pi^0}=6.29\,.
\end{equation}
Keeping $m_d=m_u$ explicitly fixed while varying $\Delta m_{\rho}$ as described above, namely through the two-body interaction kernel, leads to an increasing trend in $g_{\rho^0\pi^+\pi^-}/g_{\rho^+\pi^+\pi^0}$. This behavior persists for $m_d=2m_u$. Adopting the experimental value $\Delta m_\rho=0.7(0.8)\,\text{MeV}$, we find:
\begin{equation}
    \label{eq:RatioFinal}
    g_{\rho^0\pi^+\pi^-}/g_{\rho^+\pi^+\pi^0}=1.0024(17)(1)\,.
\end{equation}
The first uncertainty reflects the dynamical IB component, namely the isospin breaking induced through the bound-state dynamics. The second, much smaller uncertainty, accounts for the explicit variation of the current-quark mass ratio $m_d/m_u$.

In the following section, we examine the impact of IB on the $\rho$-meson decay width and, subsequently, on the HVP contribution to $a_\mu$.

\section{Effects on width difference of rho mesons and muon $g-2$}

The difference of $\rho^+-\rho^0$ meson decay widths, $\Delta\Gamma_{\rho}\equiv \Gamma_{\rho^0}-\Gamma_{\rho^+}$, is one of the parameters that, in the resonance region, controls the departure from unity (CVC prediction) of the ratio of electromagnetic $F_{\pi}^V(s)$ and weak $f_+(s)$ pion form factors \cite{Alemany:1997tn}. The individual decay widths can be extracted from the experimental resonant shapes using specific form factor models \cite{ParticleDataGroup:2026mpi}. This generally leads to different values of the  resonance parameters, in particular for $\Delta \Gamma_{\rho}$, depending of the process and the model used to extract them from data (for a recent re-analysis see for instance \cite{Davier:2025jiq}).

Alternatively, $\Delta \Gamma_{\rho}$ can be evaluated by adding up the partial widths of  exclusive $\rho^{0,+}$ decays and taking its difference \cite{Flores-Baez:2007vnd,Flores-Baez:2025nra}. We can divide the exclusive decay channels of the $\rho(770)$ meson into their (dominant) photon-inclusive two-pion decays $\pi\pi(\gamma)$ and the 'rest' (non-$\pi\pi(\gamma)$), namely $\Gamma_{\rho^i} =\Gamma(\rho^i \to \pi\pi(\gamma))+\Gamma(\rho^i \to {\rm rest})$ \cite{Flores-Baez:2007vnd,Flores-Baez:2025nra} such that
\begin{equation} \label{wdiif-1}
    \Delta \Gamma_{\rho} = \Delta \Gamma_{\rho}[\pi\pi(\gamma)]+\Delta \Gamma_{\rho}({\rm rest})\ .
    \end{equation}
According to \cite{ParticleDataGroup:2026mpi}, $\Delta \Gamma_{\rho}({\rm rest})= +0.085(17)$ MeV, and   \begin{eqnarray}\label{ppg}
 \Delta \Gamma_{\rho}[\pi\pi(\gamma)]&=& \Gamma(\rho^0\to \pi^+\pi^-) \left[ (\delta_0-\delta_+)-2\left(\frac{\delta g}{g}\right) \right.\nonumber \\ && \left. \ \ \ - \ \frac{3}{2}\left(\frac{1-\beta_0^2}{\beta_0^2}\right)\frac{\delta m_{\pi}}{m_\pi}\right]  \ . 
\end{eqnarray}
In the previous equation, $\delta_0(\delta_+)$ denotes the radiative corrections to $\rho^0\to\pi^+\pi^- (\rho^+\to \pi^+\pi^0)$ decay \cite{Flores-Baez:2025nra}, the second term is the isospin breaking in the $\rho\pi\pi$ coupling, calculated in this work (see definition in introduction), and the third term is the isospin breaking in the phase-space factors, with $\beta_0$ the velocity of pions in $\rho^0$ meson decay rest frame and $\delta m_{\pi}=m_{\pi^+}-m_{\pi^0}$. 

Using $\delta_0-\delta_+=+4.60(6)\times 10^{-3}$ \cite{Flores-Baez:2025nra} and $\delta g/g=-0.0024(17)$ calculated from Eq. (\ref{eq:RatioFinal}), we obtain for the difference of photon-inclusive $\rho\to \pi\pi$ widths
\begin{equation}
 \Delta \Gamma_{\rho}[\pi\pi(\gamma)]=+0.31(51)\ {\rm MeV}  
\end{equation}
and, taking into account the width difference of other channels as explained in Eq. (\ref{wdiif-1}), we get 
\begin{equation} \label{wdif}
 \Delta \Gamma_{\rho}=+0.39(51)\ {\rm MeV}   
\end{equation}
for the difference of total widths. This result can be compared with the corresponding average of experimental measurements  reported by the PDG  $\Delta \Gamma_{\rho}=(+0.31\pm 1.3)$ MeV \cite{ParticleDataGroup:2026mpi} and a recent data-based determination $\Delta \Gamma_{\rho}=(-0.58\pm 1.01)$ MeV \cite{Davier:2025jiq}, obtained using the Gounaris-Sakurai parametrization \cite{Gounaris:1968mw} of the pion form factors. Clearly, our result is compatible with both determinations although with a smaller error. Eventually, with a more precise measurement of the $\rho^+-\rho^0$ mass difference, the uncertainty in this theory-based  prediction of the width difference (\ref{wdif}) will be improved. 

Next, we estimate the corresponding isospin breaking correction induced by the three different sources of IB in Eq. (\ref{ppg}), into the HVP of the muon $g-2$ obtained  using tau data. The IB corrections in the HVP contribution to the anomalous magnetic moment of the muon calculated with the $\tau \to \pi\pi\nu_{\tau}(\gamma)$ data is given by (see for example \cite{Davier:2010fmf})
\begin{eqnarray} \label{diamu}
\Delta a_{\mu}^{\rm HVP, LO}[\tau, 2\pi]\!&=& \! \frac{\alpha^2m_{\tau}^2}{6|V_{ud}|^2\pi^2}\frac{{\cal B}_{\pi^pi^0}}{{\cal B}_e} \int_{4m_{\pi}^2}^{m_{\tau}^2} ds \frac{K(s)}{s} \frac{dN_{\pi\pi^0}}{N_{\pi\pi^0}ds} \nonumber \\ && \times {\cal F}(s)\left[\frac{R_{\rm IB}(s)}{S_{EW}}-1 \right] \ ,
\end{eqnarray}
 where ${\cal F}(s)=
(1-s/m_{\tau}^2)^{-2} (1+2s/m_{\tau}^2)^{-1}$ is a kinematical factor, $(1/N_{\pi\pi^0})dN_{\pi\pi^0}/ds$ is the normalized $\pi\pi^0$ spectrum in $\tau \to \pi\pi\nu_{\tau}(\gamma)$ decay, $S_{\rm EW}$ is the universal short-distance electroweak correction (including the electromagnetic correction to $\tau \to e \nu\bar{\nu}$ width, used as normalization in (\ref{diamu}))  in semileptonic tau decay. The definition of all other factors can be found in Ref. \cite{Davier:2010fmf}.

The $s$-dependent isospin breaking correction is defined  \cite{Davier:2010fmf} 

\begin{equation} \label{rib}
R_{\rm IB}(s) = \frac{{\rm FSR}(s)}{G_{\rm EM}(s)} \left(\frac{\beta_0(s)}{\beta_+(s)}\right)^3 \left|\frac{F_{\pi}^V(s)}{f_+(s)} \right|^2  
\end{equation}
where FSR$(s)$ is the final-state radiation corrrection in $e^+e^- \to \pi^+\pi^-$,  $G_{\rm EM}(s)$ is the electromagnetic long-distance correction to the two-pion decay of tau lepton, $\beta_0(s)/\beta_+(s)$ is a phase-space correction. 

The evaluation of the IB corrections that stems  from $S_{\rm EW}$, the first two factors in Eq. (\ref{rib}), and $\rho-\omega$ mixing corrections  can be found in \cite{Davier:2010fmf, Aliberti:2025beg,Davier:2023fpl, Flores-Baez:2025nra, Colangelo:2025iad,Colangelo:2025iuq,Cirigliano:2026ios}. IB corrections arising  from the last factor in Eq. (\ref{rib}) require of a  model of the pion form factors for their computation. Following refs. \cite{Davier:2010fmf,Castro:2024prg} we use the Gounaris-Sakurai (GS) \cite{Gounaris:1968mw} paramerization for the dominant $\rho(770)$ piece of the form factors to estimate the IB corrections. In addition, we use a combination of ALEPH-CLEO-OPAL-Belle data for the properly normalized spectrum $(1/N_{\pi\pi^0})dN_{\pi\pi^0}/ds$ \cite{Davier:2010fmf}. 

Thus, the difference of $\rho^+-\rho^0$  widths due to photon-inclusive $\rho\to \pi\pi$ decays produces an isospin breaking correction in the HVP contribution to the muon $g-2$, when using tau data given by

\begin{eqnarray} \label{damu}
\Delta a_{\mu}^{\rm HVP, LO}[\tau, 2\pi] &=&\left[-2.22(3), \ -2.32(1.64), \right. \nonumber \\ &&\left. \ \ \  +4.09 \right] \times 10^{-10}\ ,
\end{eqnarray}
where the three terms in the right-hand-side (r.h.s.) arise, respectively, from the corresponding terms in the r.h.s. of Eq. (\ref{ppg}). Clearly, the largest error comes from the calculation of the IB in the $\rho\pi\pi$ couplings evaluated in this work, second term in the above equation. However, it turns out to be appromately one-third of the error assigned to it in Ref. \cite{Aliberti:2025beg} by assuming and uncertainty of $O(2\alpha/\pi)$ in the IB the strong $\rho\pi\pi$ coupling. Interestingly, the sums of these three sources of the width difference $\Delta \Gamma_{\rho}$, leads to IB effects that almost cancel in $\Delta a_{\mu}^{\rm HVP, LO}[\tau, 2\pi]$.

\

\section{Isospin breaking  in the $\phi K\bar{K}$ couplings}
Once the $g_{\rho\pi\pi}$ couplings have been determined, extending our analysis to the $g_{\phi K\bar K}$ case is rather straightforward. The light-quark current masses are kept fixed at the values determined above, namely:
\begin{equation}
    m_u\leq m_q=6.78\,\text{MeV}\,,\,m_d/m_u\in[1,2]\,.
\end{equation}
For each value $m_d/m_u$, the strange-quark current quark mass $m_s$ is then fixed from the kaon sector. In particular, in the symmetric limit $m_d/m_u=1$, the value $m_s=167.2\,\text{MeV}$ yields:
\begin{equation}
\label{eq:masskaonave}
    m_{K^+}=m_{K^0}=495.03\,\text{MeV}\,.
\end{equation}
At the opposite end of the considered range, i.e. $m_d/m_u=2$, this same value of $m_s$ produces:
\begin{equation}
\label{eq:masskaonspl}
    m_{K^+}=491.66\,\text{MeV}\,,\,
    m_{K^0}=498.56\,\text{MeV}\,,
\end{equation}
whose average is consistent with the result from Eq.\,\eqref{eq:masskaonave}. Notably, despite the rather large value of the current-quark mass $m_s$, as compared with $m_{u,d}$, we find:
\begin{equation}
    M_s=530.59\,\text{MeV}\Rightarrow M_s/M_{u,d}\approx1.45\,.
\end{equation}

Note that the determination of $m_s$ is independent of the BRL parameters $\xi_{u,d,s}$, as this piece of the interaction kernel only has an impact in the vector channel. Consequently, with $m_s$ fixed from the kaon sector, $\xi_s$ can be determined by using the experimental value of $m_\phi$ as benchmark. Thus, $\xi_s^2=0.218$ entails:
\begin{equation}
    m_\phi=1019.39\,\text{MeV}\,.
\end{equation}
Once again, with the current-quark masses and $\xi_{u,d,s}$  properly determined, and having computed the corresponding meson masses, we are in a position to obtain the dressed propagators and vertices entering the evaluation of the strong couplings in Eq.\,\eqref{eq:defIA}. With the above considerations, we find:
\begin{equation} \label{phikk}
   g_{\phi K^+K^-}/ g_{\phi  K^0\bar{K}^0}=0.97(1)\,.
\end{equation}
The deviation from unity and the associated uncertainty in this ratio stems exclusively from the explicit values $m_u$, $m_d$, and $m_s$ fixed above. Since $m_u$ and $m_d$ are fixed from the pion masses, and $m_s$ is subsequently determined as a function of $m_{u,d}$ to reproduce the kaon masses, the uncertainty ultimately originates from the explicit IB, namely, from the range $m_d/m_u$ considered here. As revealed by Eqs.~\eqref{eq:masskaonave}--\eqref{eq:masskaonspl}, such a variation translates into $m_{K^0}-m_{K^+}\neq 0$ mass splitting. The absence of dynamical sources of uncertainty in $g_{\phi K^+K^-} / g_{\phi K^0\bar{K}^0}$ can be traced to the fact that none of the mesons entering the $\phi\to K\bar{K}$ process is sensitive to the $\xi_{u,d}$ parameters responsible for the dynamical mass splitting $\Delta m_\rho\neq0$. Hence, unlike the $\rho\pi\pi$ case, no analogous dynamical mechanism is present here to generate IB in this coupling ratio\,\footnote{On the other hand, because it involves the vector $K^*$ meson, the coupling $g_{K^*K\pi}$ can indeed be affected by $\xi_{u,d}$ and, consequently, depend on the dynamical mass splitting $\Delta m_\rho$. This shall be explored in a subsequent work.}.

From the ratio of $\phi \to K\bar{K}$ branching fractions, $R_{\phi}=B(\phi \to K^+K^-)/B(\phi\to K^0\bar{K^0})$, we can extract the ratio of coupling constants using
\begin{equation}
\left|\frac{g_{\phi K^+K^-}}{g_{\phi K^0\bar{K^0}}} \right| = \sqrt{\frac{R_{\phi}}{1+\delta_{KK}} \left(\frac{v_0}{v_+} \right)^3 }    \ ,
\end{equation}
where $v_+(v_0)$ denotes the charged (neutral) kaon velocity in the $\phi$ meson rest frame, and $\delta_{KK}=\delta(K^+K^-)-\delta(K^0 \bar{K}^0)=4.09\%$ is the difference of QED radiative corrections  \cite{Cremmer:1969er} including the structure-dependent effects \cite{Flores-Baez:2008owd}. 

Table\,\ref{tab1} shows the values of the ratio of $\phi \to K^+K^-/K^0 \bar{K}^0$ couplings extracted from the measured values of the corresponding branching fractions. We observe that there is a spread of values in these couplings ranging from $0.92 \sim 1.02$ which do not allow to extract a clear conclusion about the size of isospin breaking, except that they are compatible within errors  with the calculated value shown in Eq. (\ref{phikk}). Conversely, we can use the calculated IB effects from Eq. (\ref{phikk}) and predict
\begin{equation}
    R_{\phi}^{\rm Theory}=1.484\pm 0.029 \ .
\end{equation}
This result agrees with the values reported in the Particle Data Book (see Table \ref{tab1}), but it lies $2.2\sigma$ below the average reported recently by the BESIII collaboration \cite{BESIII:2025ygp,BESIII:2026lnl}.
\vspace{.5cm}
\begin{table} 
\begin{tabular}{|l|l|l|}
\hline
$R_{\phi}^{\rm exp}$ & $|g_{\phi K^+K^-}/g_{\phi K^0\bar{K^0}}|$ & Reference  \\
\hline 
$1.637\pm 0.062$ & $1.019\pm 0.019$ & \cite{BESIII:2025ygp,BESIII:2026lnl}\\
$1.470\pm 0.052$  & $0.965\pm 0.017$ & \cite{ParticleDataGroup:2026mpi} PDG(av)\\
$1.491\pm 0.032$ & $0.972\pm 0.010$ & \cite{ParticleDataGroup:2026mpi} PDG(fit) \\
\hline
\end{tabular} 
\caption{ \small Ratio of $\phi KK$ couplings extracted from  the  measured branching ratios. PDG(av/fit) denote the results obtained from the average/fit values of $R_{\phi}$ reported in  \cite{ParticleDataGroup:2026mpi}.}
\label{tab1}
\end{table}

\section{Conclusions}

Isospin breaking effects in the strong $VPP$ couplings of $V=\rho/\phi$ mesons were calculated using a symmetry-preserving Dyson-Schwinger and Bethe-Salpeter equations framework. A key feature of the approach is the inclusion of beyond-rainbow-ladder contributions in the two-body interaction kernel, providing, for the first time in these quantities, a consistent treatment of both explicit and dynamical sources of isospin breaking. This framework enables a fully nonperturbative determination of the dressed-quark propagators and Bethe-Salpeter amplitudes required for the evaluation of the strong couplings. Herein, explicit isospin breaking is introduced through the current-quark masses $m_{u,d,s}$, controlled by the ratio $m_d/m_u\in[1,2]$, while the beyond-rainbow-ladder terms in the two-body kernel provide an additional dynamical source of isospin breaking via the flavor-dependent parameters $\xi_{u,d,s}$. The former are fixed via the pion and kaon masses, while the latter control the $\rho$ meson mass-splitting and $\phi$ meson mass. 

In the $\rho\pi\pi$ sector, isospin-breaking effects are strongly suppresed, with the neutral-to-charged coupling ratio marginally exceeding unity. The associated uncertainty is also modest, below the $0.2\%$ level, and is largely driven by the dynamical sources of isospin breaking. By contrast, the $g_{\phi K^+K^-}/ g_{\phi K^0\bar{K}^0}$ ratio exhibits a more sizably breaking, at the $3\%$ level, with its uncertainty determined entirely by the explicit sources of isospin breaking. These patterns can be understood from the underlying dynamics. In the $u/d$ sector, nonperturbative mechanisms dominate, so that isospin breaking is driven primarily by the dressed-quark masses rather than by their current-quark counterparts. Extending the analysis to the strange sector, nonperturbative dynamics remain dominant; nevertheless, within the present approach, the mesons entering the $\phi K\bar{K}$ processes are insensitive to the dynamical sources of isospin breaking. This insensitivity is specific to the present treatment of this channel and would not, for instance, be expected in the $K^*K\pi$ case.

Taken together with other sources of isospin breaking in the partial widths, we obtain $\Delta \Gamma_{\rho}=+0.39(51)\ {\rm MeV}   $ for the difference  of $\rho^0-\rho^+$ total widths. This theory-based prediction agrees with results of other data-based determinations \cite{ParticleDataGroup:2026mpi,Davier:2025jiq} although it is more precise by a factor of 2$-$3. 

The Isospin Breaking in the $\rho\pi\pi$ couplings also impacts the IB corrections to the HVP of the muon $g-2$ when using the $\tau \to \pi\pi\nu_{\tau}$ spectral function \cite{Aliberti:2025beg}. When combining the results of the present work with other sources of IB corrections (radiative corrections to $\rho \to \pi\pi$  decays \cite{Flores-Baez:2025nra} and the pion mass difference) shown in Eq. (\ref{damu}), we estimate $\Delta a_{\mu}^{\rm HVP, LO}[\tau, 2\pi]=-0.25(1.64)\times 10^{-10}$ for the IB that stems from  the ratio of pion form factors in Eqs. (\ref{diamu},\ref{rib}). This result can be directly compared with the evaluation $\Delta a_{\mu}^{\rm HVP, LO}[\tau, 2\pi]=-1.62(65)\times 10^{-10}$ \cite{Davier:2010fmf} that also uses the Gounaris-Sakurai parametrization of the pion form factors. The difference between these two calculations stems from a reevaluation of the radiative corrections from Ref. \cite{Flores-Baez:2025nra} and the IB in $\rho\pi\pi$ couplings calculated in the present work. 

\section*{Acknowledgments.} We are grateful to Genaro Toledo for help with evaluation of the dispersion integral. Work supported by the Spanish Ministry of Science and Innovation (MICINN grant no.\ PID2022-140440NB-C22), and Junta de Andalucía (grant no.\ P18-FR-5057). KR acknowledges the Physics Department at Cinvestav-Zacatenco for its hospitality and support.

\bibliography{main}



\end{document}